\documentclass[conference]{IEEEtran}
\IEEEoverridecommandlockouts
\AtBeginDocument{%
  }
\usepackage{amsmath,amssymb,amsfonts}
\usepackage{graphicx}
\usepackage{textcomp}
\usepackage{xcolor}
\usepackage{url}
\usepackage{multirow}
\usepackage{framed}
\usepackage{booktabs,makecell}
\usepackage[framemethod=TikZ]{mdframed}
\usepackage{mdframed}
\usepackage{booktabs}

\newcommand{\TODO}[1]{}
\newcommand{\NOTE}[1]{}
\newcommand{\FIXME}[1]{}

\begin{document}

\title{SCOPE: Leveraging Subgoal Critiques for Code Generation}

\author{
\IEEEauthorblockN{Yueke Zhang\IEEEauthorrefmark{1}, Yifan Zhang\IEEEauthorrefmark{1}, Zihan Fang\IEEEauthorrefmark{1}, Kevin Leach\IEEEauthorrefmark{1}, Juan Zhai\IEEEauthorrefmark{2}, Wei Zhang\IEEEauthorrefmark{3}, Yu Huang\IEEEauthorrefmark{1}}
\IEEEauthorblockA{\IEEEauthorrefmark{1}Vanderbilt University\\
\{yueke.zhang, yifan.zhang.2, zihan.fang, kevin.leach, yu.huang\}@vanderbilt.edu}
\IEEEauthorblockA{\IEEEauthorrefmark{2}University of Massachusetts at Amherst\\
juanzhai@umass.edu}
\IEEEauthorblockA{\IEEEauthorrefmark{3}IBM\\
weiz@us.ibm.com}
}

%Leverage Subgoal Critiques for Code Generation

\maketitle

\begin{abstract}
Code generation with large language models (LLMs) remains unreliable because generated programs can appear correct while still violating key semantic requirements in the natural-language specification. Existing feedback-based methods improve over coder-only generation, but they often rely on unstructured critique or execution signals that do not explicitly identify what the code is semantically missing. We present \textsc{SCOPE}, a prover-initialized subgoal critic for code generation. SCOPE adapts a Lean-oriented prover model to produce three parseable feedback fields for downstream code generation: subgoals, gap analysis, and a robustness checklist. Our approach combines supervised fine-tuning, process-aligned reinforcement learning (RL), and feedback-guided inference, with two complementary rewards during RL: a dense reward for structured critique quality and a sparse reward based on whether the critique improves the coder's execution score.

Experiments show that \textsc{SCOPE} improves over the compared feedback baselines. On LiveCodeBench V6, \textsc{SCOPE} achieves 39.4\% pass@1, compared with 36.6\% for Reflexion and 20.6\% for the coder-only baseline. On BigCodeBench (Hard), it reaches 42.6\%, surpassing Reflexion at 36.5\% and coder-only generation at 34.5\%. Further analysis shows that SCOPE's gains are concentrated in tasks with concrete semantic constraints and that its localized code corrections than Reflexion's.
\end{abstract}


\section{Introduction}
%hallucination in code LLMs → tests are necessary but incomplete → LLM-generated tests are also unreliable → full formal verification is strong but costly/spec-heavy → Lean shows that machine-checkable reasoning can scale in math → can Lean-style subgoals become a practical bridge for code generation?

Large Language Models (LLMs) for coding are now mainstream~\cite{nam2024using,joel2024survey}.
AI coding assistants have moved from early trials to routine practice across industry and open source: enterprises report broad adoption and measurable productivity gains~\cite{kazemitabaar2023novices,zhao2025codinggenie}. Recent data highlights rapid growth in AI-assisted development, while field studies with large organizations document sustained usage and positive developer experience impacts~\cite{shethiya2025ai}.

However, reliability remains the central obstacle for LLM-based code generation~\cite{zhang2025practical}. Code LLMs often produce programs that look plausible but violate the user’s actual intent: they may hallucinate APIs, invent unsupported behaviors, or implement logic that passes superficial checks while failing on edge cases~\cite{chen2025empirical,wang2025llms}. Testing helps, but it does not fully solve the problem~\cite{schafer2023empirical}. In realistic software settings, developers cannot enumerate tests for every path, corner case, and semantic constraint, and recent studies show that LLM-generated tests are themselves incomplete, sometimes invalid, and often weak at exposing real defects~\cite{wang2024software,liu2024large}. Thus, passing a limited test suite is useful evidence, but it is not the same as showing that the generated code actually satisfies the intended behavior~\cite{yang2024evaluation}.

\begin{figure}[!htbp]

  \centering 
\includegraphics[width=0.5\textwidth]{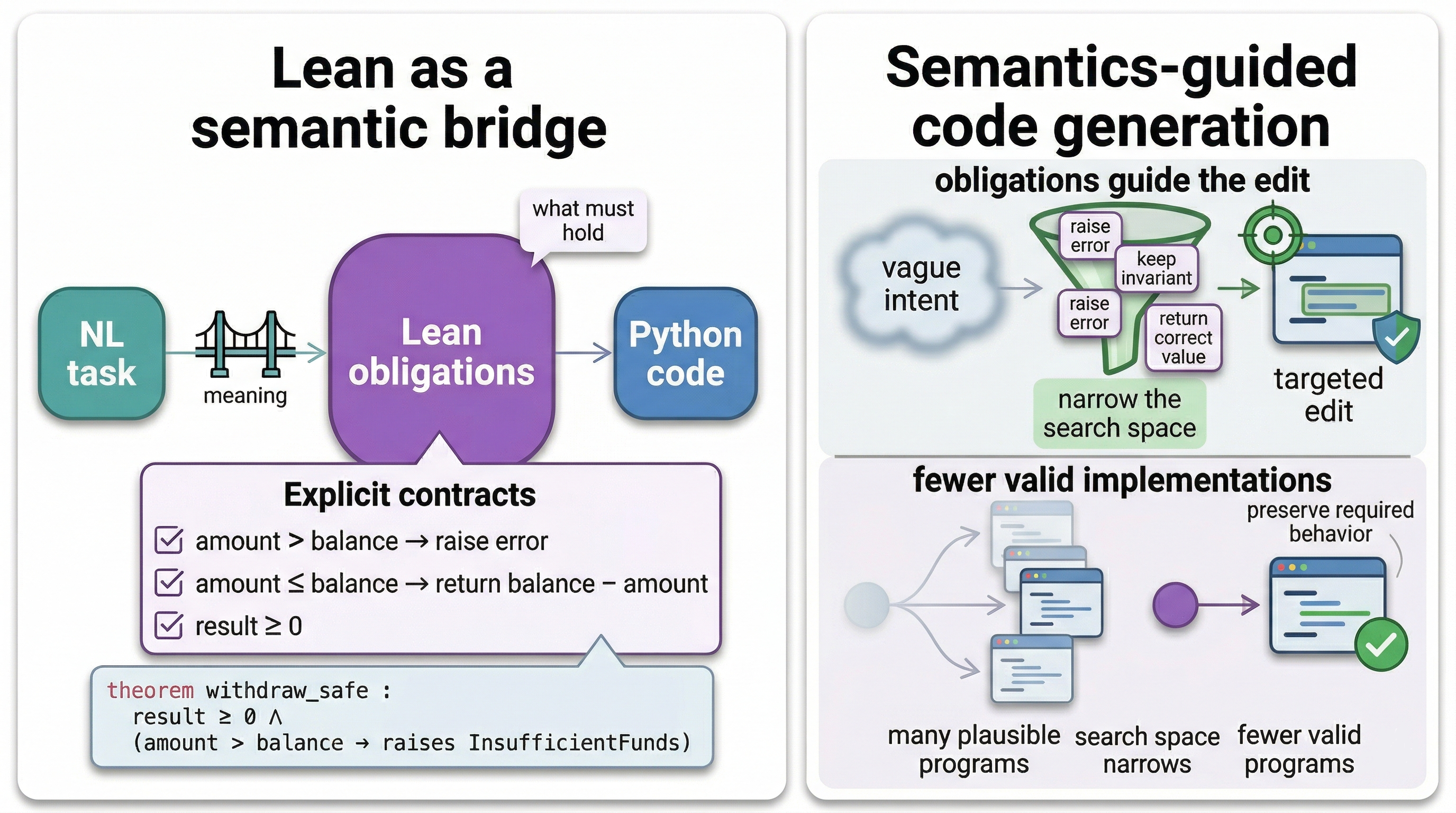}
\caption{ 
A natural-language task usually contains latent semantic requirements that are difficult to recover from execution feedback alone. SCOPE fills this gap by asking a prover-initialized critic to expose those requirements as explicit repair obligations before the coder revises the program.
}
    \label{fig:gap}

\end{figure}

A natural response to unreliable code generation is verification, but this is precisely where the difficulty deepens~\cite{first2023baldur}. 
Existing approaches based on contracts, symbolic reasoning, and SMT solvers can verify well-specified properties, yet they rely on precise formal statements of behavior that are rarely available in open-ended natural-language programming tasks~\cite{klein2009sel4,demoura2008z3}. 
Recent work at the intersection of LLMs and formal methods is encouraging: models can generate and repair proofs in interactive theorem provers and produce verifier hints for formally checked programs~\cite{first2023baldur,loughridge2024dafnybench}. 
However, these successes largely assume that the task is already formalized, that the target language is verifier-native, or that rich proof state feedback is available during search. When these assumptions are relaxed, end-to-end verified code generation remains difficult even for the most advanced LLMs ~\cite{loughridge2024dafnybench,thakur2025clever}. 
Thus, the main bottleneck is not only discharging formal obligations, but first deriving useful obligations from ambiguous natural-language intent. 
This leaves a gap between weak evidence from testing and the high cost of full manual formalization.

\begin{figure*}[!htbp]

  \centering 
\includegraphics[width=0.8\textwidth]{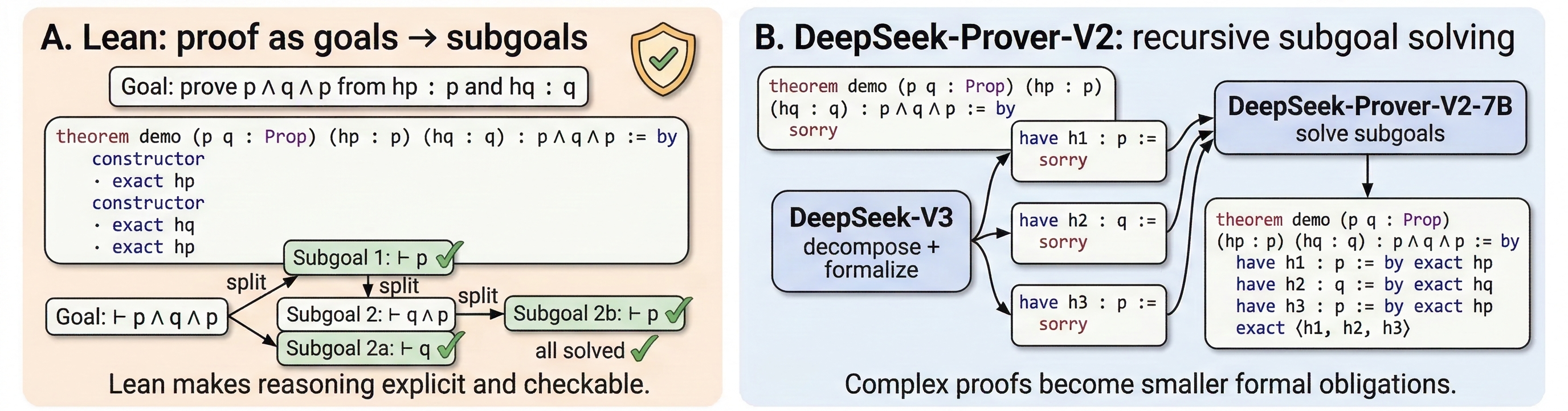}
\caption{Illustration of the subgoal-decomposition prior that motivates SCOPE. \textbf{(A)} In Lean, tactics transform one formal goal into smaller subgoals that must all be discharged. \textbf{(B)} DeepSeek-Prover-V2 is trained for Lean proof construction and inherits this bias toward decomposing a problem into intermediate obligations. SCOPE does not run this proof search on Python code; it reuses the model as a critic that emits natural-language obligations for repair.}
    \label{fig:prover_prior}

\end{figure*}

Recent progress in formal mathematics suggests a promising direction for narrowing this gap~\cite{wang2024theoremllama,zhang2025leanabell}. Proof assistants such as Lean 
provide a reasoning environment in which intermediate steps are explicit, modular, and machine-checkable under a small trusted kernel~\cite{demoura2021lean4,demoura2021lean4}. 
Moreover, the Lean ecosystem has matured substantially, with infrastructure such as LeanDojo exposing proof states, premises, and benchmarks to learning-based systems~\cite{yang2023leandojo,hubert2025olympiad}.
At the same time, Lean-oriented LLMs have shown that models can make progress by decomposing difficult problems into intermediate subgoals rather than relying on unconstrained text generation alone~\cite{ren2025deepseekproverv2,lin2025goedel}. 
These developments suggest a more limited and practical use of prover-trained models for code generation. Rather than treating Lean as a runtime verifier for Python, we ask whether the decomposition behavior learned by Lean-oriented prover models can be repurposed as structured semantic feedback for code repair~\cite{yang2025position}.

%\juan{I feel that figure 1 is way too complicated here. only some high-level components should be ok. or we can move the figure to later section like approach part.}

Motivated by this idea, we propose \textbf{SCOPE}, a prover-initialized subgoal critic for code generation. As illustrated in Figure~\ref{fig:gap}, SCOPE is designed to fill the gap between ambiguous natural-language intent and actionable repair signals. Given a task and a draft program, the critic emits three structured fields: subgoals that state what the solution should satisfy, gap analysis that compares the draft against those subgoals, and a robustness checklist that highlights boundary conditions. The output is natural language, not a machine-checked Lean proof, but it is constrained enough to be parsed and fed back to the coder. SCOPE therefore provides a practical middle ground between weak test-only feedback and expensive end-to-end formal verification. Concretely, our approach has three stages: \textbf{(1)} supervised fine-tuning to adapt a Lean-oriented prover model to the critic interface, \textbf{(2)} process-aligned reinforcement learning with dense semantic rewards and sparse execution rewards, and \textbf{(3)} feedback-based inference, where the critic's structured output guides iterative code revision.

We summarize the contributions of this paper as follows:
\begin{enumerate}
    \item We present \textbf{SCOPE}, a \emph{prover-initialized LLM} as a subgoal critic in two model code generation framework.

    \item We design a \textbf{process-aligned reward framework} for training the critic in coding tasks. It combines a \emph{dense reward} for producing structured, semantically useful critiques and a \emph{sparse reward} based on whether the critique ultimately helps the coder generate a correct program under execution based evaluation.

    \item We show that SCOPE achieves stronger performance than competitive self-refinement baselines. On LiveCodeBench~V6, SCOPE reaches 39.4\% pass@1, outperforming Reflexion at 36.6\%, Self-Refine at 33.1\%. On BigCodeBench, SCOPE achieves \textbf{42.6\%} pass@1, again surpassing Reflexion 36.5\%.

%    \item \juan{as claimed, our tool can narrow down the search space, I feel that it would be better to show how faster our tool can achieve.}

    \item We provide evidence for why SCOPE works better. Our analysis shows that, SCOPE rescues more failed solutions than Reflexion (19 vs.\ 16), achieves better bug-triggered localization (42.1\% vs.\ 31.3\% localized wins within 20 lines), and does so with more surgical repairs (28.0 vs.\ 35.0 median changed lines).
\end{enumerate}

The remainder of this paper is organized as follows. Section~II introduces the background. Section~III presents a motivating example. Section~IV reviews related work. Section~V describes the SCOPE methodology. Section~VI presents the evaluation protocol. Section~VII reports the empirical results and explains where SCOPE helps most and why. Section~VIII discusses implications and limitations. Finally, Section~IX concludes the paper.

\begin{figure*}[!htbp]

  \centering 
\includegraphics[width=0.7\textwidth]{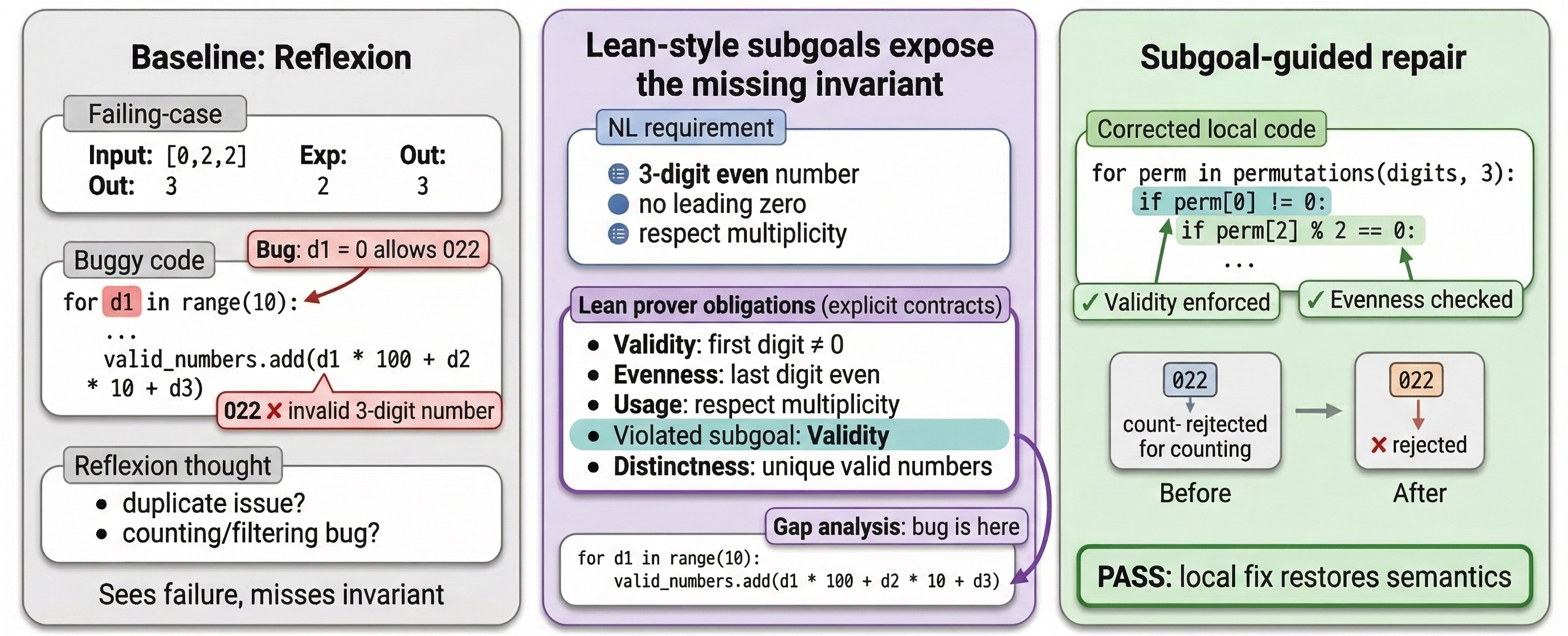}
\caption{The figure contrasts Reflexion’s broad natural-language reasoning with SCOPE’s subgoal-guided analysis, which exposes the missing invariant and maps failure to a precise local fix.}
\label{fig:comparision}
\end{figure*}
\section{Background: Why Prover-Style Decomposition Fits Code Generation}
\label{sec:background_lean}

Lean is an interactive theorem prover in which reasoning is expressed as a sequence of machine-checkable proof states. As illustrated in Figure~\ref{fig:prover_prior}A, a proof starts from a goal together with local hypotheses, and tactics transform that state by reducing one difficult goal into several smaller subgoals. A proof is complete only when all subgoals are discharged. For SCOPE, this view is an analogy and a source of model prior, not a runtime verification procedure. Each emitted subgoal is a natural-language semantic obligation, such as a precondition, an invariant, an output constraint, or an edge case requirement.

The useful property for code generation is not formal certification, but disciplined decomposition. In Lean, reasoning does not proceed as unconstrained text generation. At each step, the prover operates on an explicit proof state: a typed goal together with local hypotheses. This state rules out many irrelevant directions, and tactics further narrow the search by transforming one hard goal into a small number of concrete subgoals. SCOPE borrows this decomposition pattern for feedback: the critic should not rewrite the program or produce a long reflection, but should identify a compact set of obligations that a coder can inspect and repair.

This is exactly the kind of structure that natural-language-to-code generation lacks. A programming prompt usually bundles together several requirements, such as functional behavior, boundary conditions, interface constraints, and hidden invariants, but standard code generation leaves these requirements implicit. As a result, a code model may produce a plausible program while missing one thin semantic condition that later causes failure. SCOPE uses a prover-initialized critic as an intermediate layer because it can turn mixed natural-language intent into explicit repair obligations. Once these obligations are exposed, the search problem becomes smaller and better organized: instead of searching broadly for any revision that might pass tests, the system can focus on edits that address a specific violated condition.

DeepSeek-Prover-V2 is especially well matched to this role because it is already trained to reason through recursive subgoal decomposition rather than free-form explanation~\cite{ren2025deepseekproverv2}. SCOPE uses this model in a limited way. We do not ask it to synthesize Python, call Lean, or certify the entire program end to end. We prompt and post-train it to recover a semantic view of the task, identify which obligations the current draft appears to violate, and return those failures as structured feedback. In this sense, the prover model acts as a semantic search controller: it makes intent explicit, prunes semantically irrelevant revisions early, and converts vague failures into localized repair targets.

\section{Motivation Example}

Figure~\ref{fig:comparision} shows a real example from LiveCodeBench. The task is: \emph{count distinct 3-digit even numbers formed from a multiset of digits, allowing reuse only when multiple copies exist}. The prompt contains several constraints at once: the number must be 3-digit, the first digit cannot be zero, the last digit must be even, each digit can be used only as often as it appears in the input, and duplicate valid numbers should be counted once. Both \textsc{Reflexion}~\cite{shinn2023reflexion} and our approach are feedback-based code generation methods: they improve an initial draft by feeding critique generated by feedback LLM back to the coder for revision. The key difference is that \textsc{Reflexion} relies mainly on natural-language feedback from test traces, whereas SCOPE asks a prover-initialized critic to produce structured semantic feedback that makes missing obligations and likely failure points more explicit.

The left side of Figure~\ref{fig:comparision} illustrates this failure mode with a Reflexion setup. The coder is Qwen3-Coder-30B, and the feedback model is Qwen3-8B. The Reflexion fails because the coder generates plausible counting logic but misses one hidden semantic constraint: it allows a leading zero, so an invalid number such as 022 is counted as valid. Given the failing case $[0,2,2] \rightarrow 2$, the system knows only that the current output is wrong. It generates a long natural-language critique, but the feedback does not isolate the violated constraint. As a result, the search remains broad: the model can revise many parts of the program without directly addressing the real issue, namely that a number with a leading zero is being treated as valid.

The right side of Figure~\ref{fig:comparision} shows how SCOPE changes this process. We keep the same coder, but replace free-form feedback with a structured critic initialized from DeepSeek-Prover-V2-7B. Instead of producing broad discussion, the critic decomposes the prompt into explicit obligations and identifies that the \emph{validity} condition is violated. This turns the error into a local repair problem: the revision only needs to enforce that the first digit is non-zero. The resulting fix is small, targeted, and semantically grounded.

This example captures the main motivation for SCOPE. The advantage comes from using structured semantic feedback emitted by the critic. By converting informal intent into explicit obligations, SCOPE narrows the repair space and makes code revision more precise.

\begin{figure*}[!htbp]

  \centering 
\includegraphics[width=0.8\textwidth]{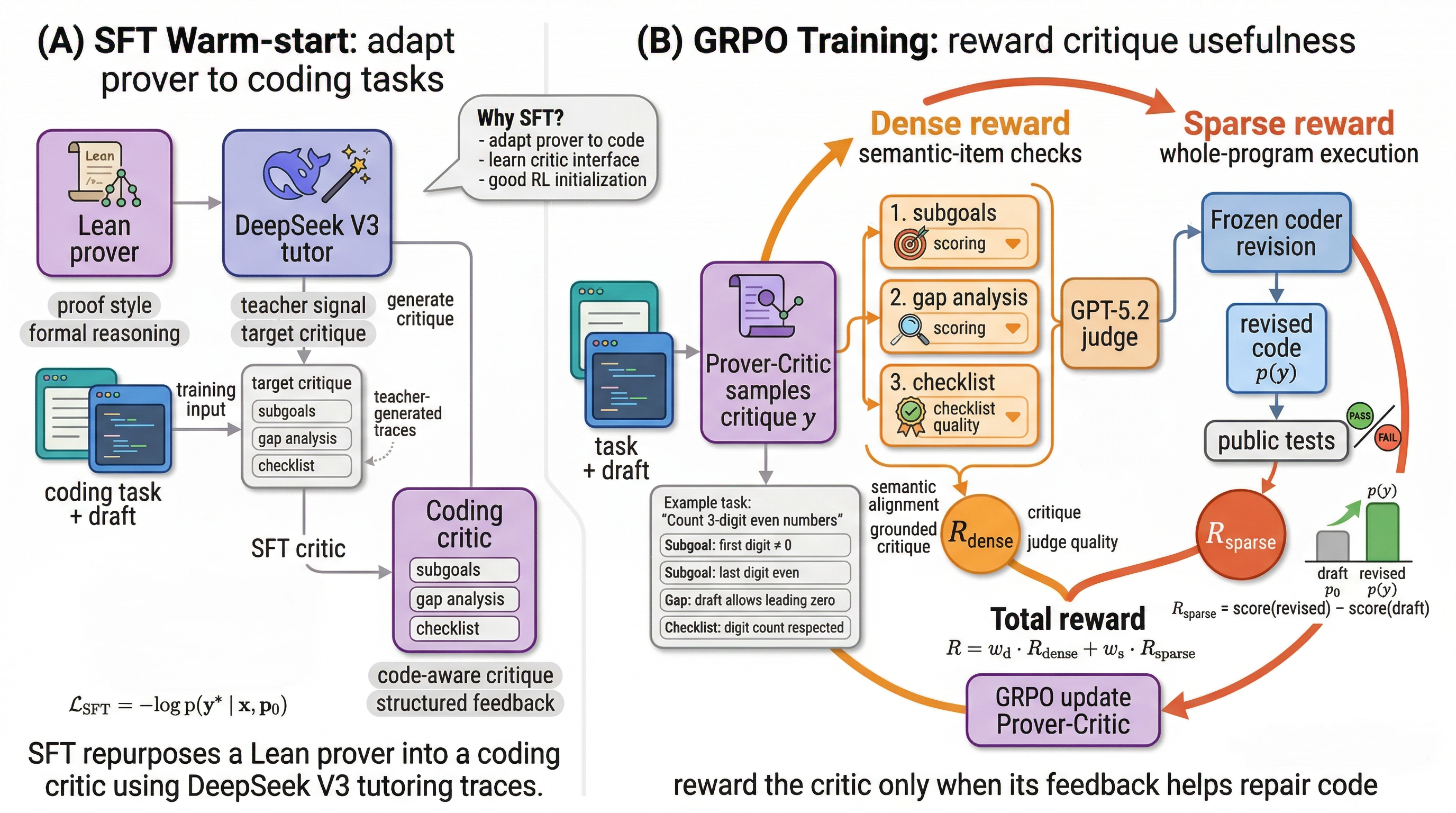}
\caption{SCOPE couples a coder with a prover-initialized critic that receives the problem and draft code, then emits structured subgoals, gap analysis, and robustness checks. During training, the critic is aligned with dense semantic rewards and sparse execution-based rewards. During inference, its critique guides iterative revision by the coder.} 

    \label{fig:meth}
\vspace{-6mm}
\end{figure*}

\section{Related Work}
\label{sec:related}
This section reviews prior work most relevant to our approach, including LLM-based code generation, feedback-driven refinement, and formal reasoning for program repair and verification.
\subsection{LLMs for Code Generation and Hallucination}

Code LLMs have rapidly advanced natural-language-to-code generation. Open models, including \textsc{Code Llama} and \textsc{StarCoder}, further improved support for multi-language generation and deployment in developer-facing tools~\cite{roziere2023codellama,li2023starcoder}.  

A growing body of work shows that the remaining reliability gap is a hallucination problem. Recent studies develop taxonomies of code hallucinations and show that these errors include invented APIs, incorrect package names, resource assumptions, mapping mistakes, and faulty program logic~\cite{liu2024hallucode,tian2024codehalu,zhang2025practical}. These failures are obvious from surface form alone: a generated program can look polished while violating hidden constraints or relying on nonexistent dependencies. Package hallucinations make this risk even more concrete by turning generation errors into potential software-supply-chain exposure~\cite{spracklen2025package}. Several mitigation strategies have therefore been proposed. For example, \textsc{De-Hallucinator} grounds generation with iteratively retrieved project-specific API context to reduce repository-level API hallucinations~\cite{eghbali2024dehallucinator}.
However, most existing approaches still ground code using retrieved context, execution feedback, or post-hoc detection, rather than explicitly representing the semantic obligations implied by the original natural-language task. 

\subsection{Feedback-Based LLM Code Generation Systems}

Feedback-based code generation systems improve on single-shot decoding by inserting an iterative critique, verification, or repair loop around a base model. A first line of work studies \emph{self-feedback}: \textsc{Self-Refine} shows that the same LLM can generate feedback on its own output and then repeatedly revise it, while \textsc{Reflexion} extends this idea with verbal reinforcement and episodic memory across attempts~\cite{madaan2023self,shinn2023reflexion}. In the coding setting, \textsc{Self-Debug} further demonstrates that execution results and natural-language explanations can help a model inspect and repair its own programs rather than merely resample candidates~\cite{chen2024selfdebug}. These methods establish an important point: even without changing model weights, test-time feedback can substantially improve code generation quality~\cite{liu2025survey,stamper2024enhancing}.

A second line of work makes the feedback signal more explicit or more structured. \textsc{CodeRL} introduces a critic trained to estimate functional correctness from unit-test outcomes, bringing feedback into the training loop rather than relying only on post-hoc prompting~\cite{le2022coderl}. More recent frameworks such as \textsc{TextGrad} cast natural-language critique as a gradient-like optimization signal over compound LLM systems~\cite{yuksekgonul2024textgrad}, while \textsc{RepairAgent} frames repair as an autonomous agentic loop that alternates among fault localization, patch generation, and validation~\cite{bouzenia2025repairagent}.

\subsection{Lean, Formal Proof, and LLM-Based Provers}

Lean is a modern interactive theorem prover with a small trusted kernel and dependent type theory, making proofs explicit and machine-checkable~\cite{demoura2021lean4}. Recent infrastructure such as \textsc{LeanDojo} has made Lean more accessible to learning-based systems by exposing proof states, premises, and benchmarks, which helped turn formal proving into a more standard LLM setting~\cite{yang2023leandojo}. Building on this ecosystem, specialized provers such as \textsc{DeepSeek-Prover} and \textsc{DeepSeek-Prover-V2} show that LLMs can learn premise use, proof search, and especially subgoal decomposition in Lean, rather than relying only on free-form reasoning~\cite{xin2024deepseekprover,ren2025deepseekproverv2}.

A related line studies LLMs for formal proof and verified programming more directly. \textsc{Baldur} explores whole-proof generation and repair with LLMs, while benchmarks such as \textsc{DafnyBench} and \textsc{CLEVER} show both the promise and the difficulty of generating machine-checked proofs or verified programs from natural-language tasks~\cite{first2023baldur,loughridge2024dafnybench,thakur2025clever}. However, most of this work assumes a verifier-native target language, a pre-existing formal specification, or direct proof-state feedback.

\section{Methodology}
Figure~\ref{fig:meth} summarizes our methodology. SCOPE has two runtime models: a frozen coder and a critic initialized from DeepSeek-Prover-V2-7B. The critic receives the natural-language problem and the coder's draft program. The methodology consists of three stages: \textbf{data construction}, where we build critique traces with a tutor LLM for code generation; \textbf{supervised fine-tuning}, where we warm-start the critic to emit the exact tag format consumed by the coder; and \textbf{process-aligned reinforcement learning}, where dense semantic rewards and sparse execution rewards further optimize the critic to produce feedback that helps the coder generate correct code.
\subsection{Data Construction}
\label{subsec:data}

We construct training data by replaying the same critique-guided revision process used by SCOPE at inference. 
To avoid overfitting to our main evaluation benchmark, we build the training set from \textbf{LiveCodeBench V1--V3}. 
Each task provides a natural-language problem statement, optional starter code, and test cases.

For each task, a coder first generates an initial draft $p_0$. 
A tutor critic then analyzes the draft and produces a structured critique $y$, including \emph{subgoals}, \emph{gap analysis}, and a \emph{checklist}. 
Conditioned on this critique, the coder produces a revised program $p_1$. 
This yields a process-level training tuple
\[
(x, p_0, y, p_1),
\]
where $x$ is the original problem.

In our pipeline, we use \textbf{Qwen3-Coder-30B} as the coder and \textbf{DeepSeek-V3} to generate critique traces. 
We store the task, draft, revision, and normalized critique fields for each example. 
We next use these critique-guided traces to warm-start the critic with supervised fine-tuning, which provides a smoother transition from data construction to the downstream RL stage. Across LiveCodeBench V1 to V3, this procedure yields 528 critique-guided training tuples. Each tuple contains the original task, the initial draft, the structured critique, and the revised program. The reference critiques are teacher-generated and therefore are not treated as ground truth proofs; they provide dense format and content supervision, while downstream execution reward later checks whether a critique actually helps repair.

\subsection{Supervised Fine-Tuning of the Prover Critic}
\label{subsec:sft}

We first warm-start the critic with supervised fine-tuning (SFT) so that it learns the subgoal decomposition interface in the code generation task. Given a problem description $x$ and an initial draft
program $p_0$, the critic is trained to generate a structured critique
\begin{equation}
    y = (s, g, c),
\end{equation}

where $s$ denotes the subgoals, $g$ the gap analysis, and $c$ the robustness checklist.
The target here is a compact semantic decomposition of the draft that can guide later repair.

We initialize from DeepSeek-Prover-V2-7B and fine-tune it on the critique traces constructed
in Section~\ref{subsec:data}. These targets are
refined by a tutor model so that the supervision is consistent in both structure and quality.
As a result, SFT serves primarily as a \emph{role-alignment} stage: it teaches the model to act as
a semantic critic in the code generation task instead of mathematical proving.

The supervised completion is the exact format used in the implementation:
\begin{quote}
\small
\texttt{<subgoal>} $s$ \texttt{</subgoal>}
\texttt{<gap\_analysis>} $g$ \texttt{</gap\_analysis>}
\texttt{<checklist>} $c$ \texttt{</checklist>}.
\end{quote}

Formally, for each training example $(x, p_0, y^\star)$, where
\begin{equation}
y^\star = (s^\star, g^\star, c^\star),
\end{equation}
we optimize the standard conditional language-model objective
\begin{equation}
\mathcal{L}_{\mathrm{SFT}}(\theta)
=
-\sum_{t=1}^{|y^\star|}
\log p_\theta\!\left(y^\star_t \mid x, p_0, y^\star_{<t}\right).
\end{equation}

In implementation, the prompt tokens corresponding to $(x,p_0)$ are masked out, so the loss is
applied only to the completion tokens of the structured critique.
Intuitively, SFT teaches the model
\emph{what kind of critique to produce}, while the later RL stage teaches it
\emph{which critiques actually help the coder}. In this sense, SFT provides the structural prior
that makes process-aligned reinforcement learning feasible.

\subsection{Process-Aligned Reinforcement Learning with Dense and Sparse Rewards}
\label{subsec:rl}

SFT alone does not guarantee that a critique is useful for downstream repair. A critic may produce
well-formed subgoals and fluent gap analysis that look reasonable to a human reader, yet still fail
to help the coder generate a correct program. We therefore further optimize the critic with Group Relative Policy Optimization (GRPO),
using a reward that combines \emph{dense semantic guidance} with \emph{sparse execution-based feedback}.

For a task $x$ and draft program $p_0$, the critic policy $\pi_\theta$ samples a structured critique
$y$. This critique is then fed to the coder, which produces a revised program $p(y)$.
The revised program is executed on benchmark tests, and the critic is rewarded according to whether
its critique actually improves downstream correctness. This makes the RL signal \emph{process-aligned}:
The critic is rewarded not for writing code, but for producing feedback that helps another model repair code. 

We optimize the policy with GRPO:
\begin{equation}
    \max_{\theta}\;
\mathbb{E}_{y \sim \pi_\theta(\cdot \mid x,p_0)}
\left[ R(y) \right]
-
\beta \,
\mathrm{KL}\!\left(
\pi_\theta(\cdot \mid x,p_0)\;\|\;\pi_{\mathrm{ref}}(\cdot \mid x,p_0)
\right),
\end{equation}

where $\pi_{\mathrm{ref}}$ is the reference policy initialized from SFT, and $\beta$ controls
the strength of KL regularization.

\paragraph{Dense reward.}
The dense reward encourages the generated critique to remain structurally valid, semantically aligned,
and practically readable. We define
\begin{equation}
   R_{\mathrm{dense}}(y)
=
\lambda_{\mathrm{emb}}\,R_{\mathrm{emb}}(y)
+
\lambda_{\mathrm{qual}}\,R_{\mathrm{qual}}(y). 
\end{equation}

The first term, $R_{\mathrm{emb}}$, measures semantic alignment between the generated critique and the
teacher-generated targets in the training trace. Concretely, it scores the generated subgoals, gap analysis,
and checklist against their target counterparts:
\begin{equation}
   R_{\mathrm{emb}}(y)
=
\alpha_s\,\mathrm{sim}(s,s^\star)
+
\alpha_g\,\mathrm{sim}(g,g^\star)
+
\alpha_c\,\mathrm{sim}(c,c^\star), 
\end{equation}
where $\mathrm{sim}()$ denotes an embedding-based similarity function and
$\alpha_s+\alpha_g+\alpha_c=1$.

The second term, $R_{\mathrm{qual}}$, scores critique quality. Its role is to reward outputs that are
concise, grounded in the current draft, parseable by the revision prompt, and structurally useful for repair. In our main setting,
this signal is provided by a frozen GPT-5.2 judge, which evaluates whether the critique is actionable
rather than merely fluent. The semantic similarity alone is not
enough: a critique can overlap with the target text yet still be vague or unhelpful. 

\paragraph{Sparse reward.}
The sparse reward captures the actual downstream value of the critique from testing. After sampling $y$, we invoke
the coder to revise $p_0$ into $p(y)$, then run the revised program on benchmark tests. Let
$S(\cdot)$ denote the shaped execution score, which reflects not only binary pass/fail but also
partial test progress and lightweight penalties for runtime failures. We then define the sparse reward as
\begin{equation}
    R_{\mathrm{sparse}}(y)
=
S\!\left(p(y)\right) - S(p_0).
\end{equation}

It rewards the critic only for \emph{improvement over the original
draft}. As a result, the critic is encouraged to generate critiques
that expose the draft's missing semantic obligations and help the coder make targeted repairs.

\paragraph{Final reward.}
Our final reward combines the dense and sparse terms:
\begin{equation}
    R(y)
=
w_{\mathrm{dense}}\,R_{\mathrm{dense}}(y)
+
w_{\mathrm{sparse}}\,R_{\mathrm{sparse}}(y),
\end{equation}

This combination reflects the two goals of SCOPE. The dense term keeps the critic anchored to the intended
semantic structure and prevents drift into unstructured free-form text. The sparse term ensures that the
critic is ultimately optimized for what matters most: whether its feedback helps the coder produce a correct
program under execution-based evaluation.

A useful way to interpret the two terms is as follows. The dense reward teaches the critic to produce
\emph{well-formed critiques}; the sparse reward teaches it to produce \emph{useful critiques}. SCOPE needs both.
Without the dense term, optimization becomes unstable and the model may exploit noisy execution signals.
Without the sparse term, the critic may learn to imitate well-formed obligations that are linguistically
plausible but do not improve code generation in practice.

\subsection{Inference: SCOPE as a Critique-Guided Revision Loop}
\label{subsec:inference}

At inference time, SCOPE instantiates a two-model revision loop between a coder and a
post-trained critic. Given a task $x$, the coder first produces an initial draft
program $p_0$. The critic then receives $(x,p_0)$ and returns a structured critique
\begin{equation}
y = (s, g, c),
\end{equation}

where $s$, $g$, and $c$ denote the subgoals, gap analysis, and robustness checklist,
respectively. Conditioned on this critique, the coder produces a revised program
\begin{equation}
    p_1 = C(x, p_0, y),
\end{equation}

where $C(\cdot)$ denotes the frozen coder. In this way, inference mirrors the same
critic-guided repair process used during training, ensuring that the critic is deployed
in exactly the role for which it was optimized.
The subgoals define what the program must satisfy, the gap analysis explains
where the current draft violates those obligations, and the checklist highlights boundary
conditions that are easy to miss during revision. By packaging these three components
into a single output, SCOPE gives the coder a unified repair target rather than a set of
loosely related suggestions.

Formally, the inference objective is not to search directly over programs, but to improve
the draft through critic-guided revision. If $S(\cdot)$ denotes the hidden execution-based
correctness signal, SCOPE aims to produce critiques such that
\begin{equation}
    S(p_1) > S(p_0),
\end{equation}

and, more generally, such that the revised program moves closer to satisfying the
problem's latent semantic constraints. This formulation reflects the central intuition of
our method: the critic adds value not by replacing the coder, but by exposing the
obligations that the current draft has not yet made explicit.

\section{Evaluation}
\label{sec:method}
We organize the evaluation around three questions:
\\RQ1: Does SCOPE improve end-to-end code generation accuracy over coder-only and iterative baselines? 
\\RQ2: Where are these gains concentrated: are they consistent across benchmarks, or do they appear mainly in particular difficulty levels and problem categories? 
\\RQ3: When SCOPE improves a failed baseline solution, does it do so through more precise repair behavior, such as localized edits or specific types of semantic corrections?
\subsection{Baselines}

\paragraph{Coder-Only}

This is the coder-only baseline.\textbf{Qwen3-Coder-30B} receives the problem description and function signature, generates a single solution, and that solution is evaluated directly. There is no critic model and no refinement step.

\paragraph{Self-Refine~\cite{madaan2023self}}

This baseline keeps \textbf{Qwen3-Coder-30B} as the coder and uses \textbf{Qwen3-8B} to generate a lightweight natural language critique of the draft. The feedback points out weaknesses in the current solution and suggests possible improvements in text, after which the coder revises the solution once based on that critique. This baseline tests whether generic textual self feedback is sufficient for code repair.

\paragraph{Reflexion~\cite{shinn2023reflexion}}

This baseline also uses \textbf{Qwen3-Coder-30B} as the coder and \textbf{Qwen3-8B} as the feedback model. Unlike Self-Refine, the critic additionally observes execution feedback, such as failing inputs, expected outputs, and runtime errors. It then produces natural language debugging feedback, which the coder uses for iterative revision. This setting represents a strong execution guided interactive baseline.

\paragraph{SCOPE (Untrained)}

This variant uses \textbf{Qwen3-Coder-30B} as the coder and a raw \textbf{DeepSeek-Prover-V2-7B} model as the critic. The critic is prompted to produce structured subgoals, gap analysis, and a checklist, but it is used without SFT or RL. This comparison isolates the value of the prover-initialized critic before adaptation.

\paragraph{SCOPE (SFT)}

This variant keeps the same coder, \textbf{Qwen3-Coder-30B}, and replaces the critic with an \textbf{SFT-adapted DeepSeek-Prover-V2-7B}. The goal of this comparison is to measure how much role alignment alone helps, before reinforcement learning is introduced.

\paragraph{SCOPE (Full).}
Our full system again uses \textbf{Qwen3-Coder-30B} as the coder, but the critic is the fully post-trained \textbf{DeepSeek-Prover-V2-7B} after both SFT and GRPO. The critic outputs structured semantic feedback in the form of subgoals, gap analysis, and a robustness checklist, which the coder uses for revision.

\subsection{Benchmarks, Metrics, and Analysis Protocol}
\label{sec:eval_protocol}

\paragraph{Benchmarks.}
We evaluate primarily on Python code generation with LiveCodeBench V6, using the official execution harness. This subset contains 175 tasks, each defined by a natural-language problem description together with a Python function signature. Our setting is a controlled same-coder repair comparison. To test whether the method generalizes beyond the benchmark used most heavily during development, we also evaluate on BigCodeBench-Complete (Hard), which contains 148 Python tasks. 

\paragraph{Primary metric.}
Our main metric is pass@1, defined as the fraction of tasks for which a method produces a final submission that passes all hidden tests. For the coder-only baseline, pass@1 is computed from its single generated solution.

\paragraph{Pairwise comparison and failure modes.}
Beyond overall pass@1, we compare SCOPE against each baseline on a per-task basis. For each pair, we count \emph{wins}, where SCOPE solves a task that the baseline fails, and \emph{regressions}, where the baseline solves a task that SCOPE fails. This analysis shows whether SCOPE's gains come from broad improvements across many tasks or only from a small number of isolated cases. For unsolved tasks, we also group failures into four categories: \emph{crash-like failures}, which include non-\texttt{AssertionError} runtime failures; \emph{assertion-only failures}, where the program runs but does not satisfy the hidden tests; \emph{mixed failures}, where multiple failure types appear across attempts. This lets us examine not only whether SCOPE improves final accuracy, but also whether it changes the kinds of errors the system tends to make.

\paragraph{Difficulty and category breakdown.}
To understand where the method helps most on LiveCodeBench, we join each task's official difficulty label with the corresponding binary pass/fail outcome and aggregate solved counts by difficulty level. We report results for Easy, Medium, and Hard tasks separately. We also analyze performance by algorithmic category, following the benchmark's LeetCode style taxonomy, such as array, string, hash table, math, greedy, dynamic programming, graph, tree, binary search, and simulation. This breakdown helps show whether the gains are broadly distributed or concentrated in particular classes of problems.

\paragraph{Repair localization and revision analysis.}
We also study where successful repairs occur and what kinds of revisions they make. For tasks where a coder-only approach fails but a target method succeeds, we perform a manual fault-localization check on the failing coder-only solution. Concretely, we inspect the failing program together with its execution feedback, including traceback information, assertion messages, and failing input \& output behavior, and then mark the line, or smallest contiguous region, most directly responsible for the failure. We refer to the center of this marked region as the trigger line, denoted by $\ell_{\text{trigger}}$. We then compare the failing coder-only solution against the final successful solution and identify the main edited hunk; the center of that hunk is treated as the fix line, denoted by $\ell_{\text{fix}}$. We summarize localization quality by the absolute line distance
\begin{equation}
    d = \left| \ell_{\text{fix}} - \ell_{\text{trigger}} \right|.
\end{equation}

Smaller $d$ indicates that the successful repair stays closer to the inferred fault location.
\begin{table*}[t]
\centering
\small
\caption{Pass@1 by \textbf{Difficulty} on LiveCodeBench V6 across six approaches. Best per row in \textbf{bold}.}
\label{tab:difficulty}
\begin{tabular}{lrrrrrr}
\toprule
\textbf{Difficulty} & \textbf{Coder-Only} & \textbf{Reflexion} & \textbf{Self-Refine} & \textbf{SCOPE (Untrained)} & \textbf{SCOPE (SFT)} & \textbf{SCOPE (Full)} \\
\midrule
Easy    & 26/43 (60.5\%) & 39/43 (90.7\%) & 39/43 (90.7\%) & 41/43 (95.3\%) & 41/43 (95.3\%) & \textbf{42/43 (97.7\%)} \\
Medium  &  8/52 (15.4\%) & 16/52 (30.8\%) & 16/52 (30.8\%) & 17/52 (32.7\%) & 16/52 (30.8\%) & \textbf{18/52 (34.6\%)} \\
Hard    &  2/80 ( 2.5\%) & \textbf{9/80 (11.2\%)} &  3/80 ( 3.8\%) &  6/80 ( 7.5\%) &  5/80 ( 6.2\%) & \textbf{9/80 (11.2\%)} \\
\midrule
Overall & 36/175 (20.6\%) & 64/175 (36.6\%) & 58/175 (33.1\%) & 64/175 (36.6\%) & 62/175 (35.4\%) & \textbf{69/175 (39.4\%)} \\
\bottomrule
\end{tabular}
\end{table*}
To quantify edit magnitude, we compute the changed-line count between the failing solution and the final successful solution. Let $L_{\text{base}}$ and $L_{\text{final}}$ be the sets of non-empty lines in the two versions after line based alignment. This yields a directly comparable trigger--fix view across methods under the same failure pool.

%For the BigCodeBench cross-method repair analysis, we use a shared coder-only anchor so that all compared methods are evaluated against the same failure source. Specifically, for each task where the coder-only solution does not pass and a target method does pass, we treat the coder-only traceback or failing assertion as the bug signal, infer a trigger line in the coder-only code, diff the coder-only final code against the target method's final code, and define the target fix line as the center of the largest changed diff hunk. This yields a directly comparable trigger--fix view across methods under the same failure pool.

Under this protocol, we report several complementary repair metrics. \textbf{N Wins} is the number of tasks where the coder-only solution fails but the target method succeeds. \textbf{Recovery Rate} normalizes this count by the size of the coder-only non-pass pool. 
\textbf{Localized Wins @20} is the fraction of all recovered tasks whose fix lies within 20 lines of the trigger, that is, $|\ell_{\text{fix}}-\ell_{\text{trigger}}| \le 20$. 
%\textbf{Trigger+Fix Coverage} measures how often localization can be evaluated at all, because both the bug trigger and the repair location are identifiable.
%\textbf{Localized Share Among Covered} applies the same locality criterion only to the subset where both trigger and fix are observable. 
\textbf{Trigger+Fix Coverage} is the fraction of recovered tasks for which localization can be evaluated under our protocol. A task is counted as covered only when we can identify both (1) a trigger line in the coder-only failing solution from its failure signal, and (2) a dominant fix line in the target successful solution from the main diff hunk.
\textbf{Localized Share Among Covered} is the fraction of these covered tasks whose fix lies within 20 lines of the trigger. This metric conditions on the subset where a trigger-to-fix comparison is actually well defined.
\textbf{Median Changed Lines} is the median line-diff size between the coder-only solution and the target final solution. \textbf{Median Changed Ratio} further normalizes edit size by the length of the coder-only program. Higher is better for recovery and localization metrics, while lower is better for edit-size metrics.

\subsection{Training settings}
We train the critic in two stages: a supervised warm-start followed by GRPO-based reinforcement learning. 
All training is conducted on 2$\times$ NVIDIA RTX A6000 GPUs using parameter-efficient fine-tuning. 
For both stages, we adopt \textbf{QLoRA} with rank $r=16$, $\alpha=32$, dropout $0.05$, all-linear target modules, and 4-bit NF4 quantization. 
In the SFT stage, we fine-tune \textbf{DeepSeek-Prover-V2-7B} with optimizer AdamW, gradient accumulation 8, learning rate $1\times 10^{-4}$, warmup ratio $0.03$, and 80 training steps. 
We then initialize GRPO from the SFT checkpoint and continue training with the same backbone and LoRA setting for 120 maximum steps. 
The RL training mixture keeps broad regenerated LiveCodeBench V1--V3 data and upweights focused hard-fail slices used for curriculum. We therefore interpret the LiveCodeBench V6 result as an in-family repair evaluation and use BigCodeBench as the cleaner cross-benchmark generalization check.

\section{Results}
\label{sec:results}
We organize the results from overall effectiveness to deeper analysis: we first compare SCOPE against baselines on benchmark performance, then study behavior across difficulty levels and problem categories, and finally examine where the gains come from through a fine-grained error analysis.

\begin{table*}[t]
\centering
\small
\caption{Pass@1 by \textbf{Algorithmic Category} on LiveCodeBench V6. Best per category column in \textbf{bold}.}
\label{tab:category}
\begin{tabular}{lrrrrrrr}
\toprule
\textbf{Approach} & \textbf{Array} & \textbf{DP} & \textbf{Graph} & \textbf{Hash Map} & \textbf{Math} & \textbf{Simulation} & \textbf{String} \\
\midrule
Coder-only          & 0.000 & 0.000 & 0.063 & 0.385 & 0.262 & 0.214 & 0.333 \\
Reflexion         & \textbf{0.455} & 0.167 & \textbf{0.375} & \textbf{0.692} & 0.381 & 0.286 & \textbf{0.500} \\
Self-Refine       & \textbf{0.455} & 0.000 & 0.313 & 0.615 & 0.333 & 0.214 & \textbf{0.500} \\
SCOPE (Untrained)   & \textbf{0.455} & \textbf{0.250} & 0.250 & \textbf{0.692} & \textbf{0.405} & 0.214 & \textbf{0.500} \\
SCOPE (SFT)         & \textbf{0.455} & 0.167 & 0.250 & 0.615 & 0.381 & 0.286 & 0.444 \\
SCOPE (Full)              & \textbf{0.455} & \textbf{0.250} & \textbf{0.375} & \textbf{0.692} & \textbf{0.405} & \textbf{0.357} & 0.444 \\
\bottomrule
\end{tabular}
\end{table*}

% Requires: \usepackage{booktabs}
\begin{table*}[t]
\centering
\small
\caption{Comparison of coder-only, Reflexion, SCOPE-SFT, and SCOPE on BigCodeBench. 
Solved and pass@1 report overall performance. 
``SCOPE Wins'' and ``SCOPE Regr.'' are measured relative to SCOPE (ours). 
Failure categories summarize the remaining unsolved cases.}
\label{tab:bigcodebench_overall}
\begin{tabular}{lrrrrrrrr}
\toprule
\textbf{Method} & \textbf{Solved} & \textbf{pass@1} & \textbf{SCOPE Wins} & \textbf{SCOPE Regr.} & \textbf{Fail-Crash} & \textbf{Fail-AssertionOnly} & \textbf{Fail-Mixed}  \\
\midrule
Coder-only & 51/148 & 0.3446 & 19 & 7 & 30 & 53 & 14  \\
Reflexion             & 54/148 & 0.3649 & 16 & 7 & 33 & 50 & 10  \\
SCOPE (SFT)             & 62/148 & 0.4189 &  4 & 3 & 23 & 51 & 11  \\
SCOPE (Full)          & 63/148 & 0.4257 & -- & -- & 21 & 52 & 11  \\
\bottomrule
\end{tabular}
\end{table*}

\begin{table*}[t]
\centering
\small
\setlength{\tabcolsep}{4pt}
\caption{Coder-only-anchored localization comparison on BigCodeBench-Complete (Hard). 
Each row uses the same coder-only coder-only baseline as the anchor. 
We consider tasks where the coder-only solution fails and the target method succeeds. 
Higher is better for recovery and localization metrics; lower is better for edit-size metrics. 
Best values are shown in bold.}
\label{tab:bigcodebench_localization}
\begin{tabular}{lccccccc}
\toprule
\makecell{\textbf{coder-only} \\ \textbf{$\rightarrow$ Target}}
& \makecell{\textbf{N} \\ \textbf{Wins}}
& \makecell{\textbf{Recovery} \\ \textbf{Rate}}
& \makecell{\textbf{Trigger+Fix} \\ \textbf{Coverage}}
& \makecell{\textbf{Localized} \\ \textbf{Wins @20}}
& \makecell{\textbf{Localized Share} \\ \textbf{Among Covered}}
& \makecell{\textbf{Median Changed} \\ \textbf{Lines} $\downarrow$}
& \makecell{\textbf{Median Changed} \\ \textbf{Ratio} $\downarrow$} \\
\midrule
Coder-only $\rightarrow$ Reflexion
& 16
& 16.5\%
& 37.5\%
& 31.3\%
& 83.3\%
& 35.0
& 0.703 \\
Coder-only $\rightarrow$ SCOPE
& \textbf{19}
& \textbf{19.6\%}
& \textbf{47.4\%}
& \textbf{42.1\%}
& \textbf{88.9\%}
& \textbf{28.0}
& \textbf{0.577} \\
\bottomrule
\end{tabular}
\end{table*}

\subsection{RQ1: Does SCOPE Improve Overall Code Generation Accuracy Across Benchmarks?}
\label{sec:rq1_analysis}

In table~\ref{tab:difficulty}, the comparison isolates the effect of the \emph{feedback mechanism}: no feedback for the Coder-Only, unconstrained self-critique for Self-Refine, execution trace guided natural language feedback for Reflexion, and prover-initialized subgoal feedback for SCOPE. Under this setup, SCOPE (Full) is the best-performing system overall. 
On LiveCodeBench V6, SCOPE (Full) solves \textbf{69/175} tasks, achieving \textbf{39.4\% pass@1}, compared with \textbf{64/175} (\textbf{36.6\%}) for Reflexion, \textbf{58/175} (\textbf{33.1\%}) for Self-Refine, \textbf{64/175} (\textbf{36.6\%}) for SCOPE (Untrained), \textbf{62/175} (\textbf{35.4\%}) for SCOPE (SFT), and \textbf{36/175} (\textbf{20.6\%}) for the coder-only baseline. Relative to the strongest baseline, Reflexion, SCOPE improves by \textbf{5 additional solved tasks} and \textbf{+2.8} absolute pass@1 points; relative to the coder-only baseline, the gain is \textbf{33 additional solved tasks} and \textbf{+18.8} points.

We also evaluate on BigCodeBench (Hard) as a cross-benchmark generalization check. This second benchmark differs in composition and is evaluated with its own official calibrated execution setting. The result is consistent with the main benchmark in table~\ref{tab:bigcodebench_overall}: SCOPE (Full) achieves \textbf{63/148} solved tasks (\textbf{42.6\% pass@1}), compared with \textbf{54/148} (\textbf{36.5\%}) for Reflexion, \textbf{51/148} (\textbf{34.5\%}) for coder-only generation, and \textbf{62/148} (\textbf{41.9\%}) for SCOPE (SFT). SCOPE remains the strongest method even when evaluated on a second execution-based benchmark with a different task distribution.

Overall, SCOPE delivers the strongest performance in our main evaluation, and this advantage remains visible when we test on a second benchmark designed to reduce concern about benchmark-specific fitting.

\subsection{RQ2: On Which Kinds of Tasks Does SCOPE Help Most, by Difficulty and Problem Category?}

In Table~\ref{tab:difficulty}, we demonstrate the difficulty breakdown performance. On \textbf{Easy} tasks, SCOPE (Full) achieves the best result with \textbf{42/43} solved (\textbf{97.7\%}), improving over Reflexion and Self-Refine, both at \textbf{39/43} (\textbf{90.7\%}), and over the coder-only baseline at \textbf{26/43} (\textbf{60.5\%}). On \textbf{Medium} tasks, SCOPE again obtains the strongest result with \textbf{18/52} solved (\textbf{34.6\%}), compared with \textbf{16/52} (\textbf{30.8\%}) for Reflexion and Self-Refine and \textbf{8/52} (\textbf{15.4\%}) for the baseline. On \textbf{Hard} tasks, SCOPE reaches \textbf{9/80} (\textbf{11.2\%}), matching Reflexion and clearly outperforming Self-Refine at \textbf{3/80} (\textbf{3.8\%}) and the baseline at \textbf{2/80} (\textbf{2.5\%}).
This pattern suggests that SCOPE consistently improves reliability up to the medium range and remains strong on hard tasks, while more exploratory feedback can still sometimes rescue especially difficult cases.

In Table~\ref{tab:category},
the category breakdown clarifies \emph{where} SCOPE's gains come from. SCOPE (Full) is the strongest in most of the reported categories. On \textbf{array} tasks, the baseline solves none of the tasks (\textbf{0.000}), whereas all feedback-based methods, including SCOPE, reach \textbf{0.455}. On \textbf{dynamic programming}, SCOPE reaches \textbf{0.250}, exceeding Reflexion at \textbf{0.167} and both the baseline and Self-Refine at \textbf{0.000}. It fits the intuition that dynamic programming requires preserving intermediate invariants and state relationships that benefit from explicit subgoal decomposition.

SCOPE is also strong on categories where hidden constraints are global rather than purely local. On \textbf{hash-map} tasks, SCOPE reaches \textbf{0.692}, tying Reflexion for the best result and improving over the baseline at \textbf{0.385}. On \textbf{math} tasks, SCOPE achieves the best score at \textbf{0.405}, compared with \textbf{0.381} for Reflexion and \textbf{0.262} for the baseline. On \textbf{simulation}, SCOPE attains \textbf{0.357}, again the highest value in the table, improving over Reflexion at \textbf{0.286} and the baseline at \textbf{0.214}. These gains are consistent with the intended role of SCOPE: making latent semantic requirements explicit enough that the coder can repair them systematically rather than relying only on free-form critique.
On \textbf{graph} problems, SCOPE reaches \textbf{0.375}, tying Reflexion for the best result and outperforming Self-Refine at \textbf{0.313} and the baseline at \textbf{0.063}. Graph problems are often structurally complex. They can also admit multiple plausible solution paths. As a result, SCOPE’s advantage is smaller on these tasks. Success often depends more on exploring alternative strategies than on correcting a specific semantic error.

SCOPE improves over Reflexion on dynamic programming (\textbf{0.250} vs.\ \textbf{0.167}), math (\textbf{0.405} vs.\ \textbf{0.381}), and simulation (\textbf{0.357} vs.\ \textbf{0.286}), while tying Reflexion on graph and hash-map tasks. These are categories where failures often come from violating a recurrence, numeric invariant, uniqueness condition, or state-transition rule. The gains are smaller where the main challenge is broad strategy exploration rather than fixing a specific violated obligation.  SCOPE is most useful when a draft is close enough to repair and the missing behavior can be expressed as a small number of explicit obligations.

\subsection{RQ3: What Explains SCOPE’s Advantage Over Baselines in Localization and Revision Behavior?}

Table~\ref{tab:bigcodebench_overall} combines three complementary views of performance: overall pass@1, pairwise win/loss decomposition against each baseline, and failure-mode shift on the remaining unsolved tasks. Under this view, SCOPE achieves the strongest overall result on BigCodeBench-Complete (Hard), solving \textbf{63/148} tasks (\textbf{42.6\% pass@1}), compared with \textbf{54/148} (\textbf{36.5\%}) for Reflexion and \textbf{51/148} (\textbf{34.5\%}) for coder-only generation. Relative to Reflexion, this corresponds to a net gain of \textbf{+9} tasks (\textbf{16} SCOPE wins vs.\ \textbf{7} regressions); relative to coder-only generation, the gain is \textbf{+12} tasks (\textbf{19} wins vs.\ \textbf{7} regressions).
The same table also shows that SCOPE produces the fewest crash-like failures among the compared methods. In particular, \textbf{Fail-Crash} cases decrease from \textbf{33} under Reflexion to \textbf{21} under SCOPE, and from \textbf{30} under coder-only generation to \textbf{21}. Crash failures are qualitatively different from pure assertion failures: they often indicate unresolved semantic mismatches, such as incorrect exception behavior, wrong API usage, or failure to satisfy hidden structural constraints in the harness. SCOPE is more often repairing the semantic conditions that prevent the program from executing correctly in the first place.

Table~\ref{tab:bigcodebench_localization} demonstrates the localization comparison on the same coder-only baseline for both Reflexion and SCOPE. Under this shared baseline, SCOPE rescues more coder-only failures than Reflexion, improving from 16 wins to 19 wins, or from 16.5\% to 19.6\% recovery on the 97 coder-only non-pass tasks. SCOPE also yields higher trigger+fix coverage (47.4\% vs.\ 37.5\%), which means a larger fraction of its successful repairs can be aligned to a concrete bug trigger and a concrete fix location.
The difference also appears in localized rescue behavior. SCOPE achieves 8 localized wins within a 20-line neighborhood of the coder-only trigger, compared with 5 for Reflexion; as a rate over all wins, this is 42.1\% for SCOPE versus 31.3\% for Reflexion. Even after restricting to cases where both trigger and fix can be identified, SCOPE remains better (88.9\% vs.\ 83.3\%). At the same time, SCOPE is more surgical: its median changed-line count is smaller (28.0 vs.\ 35.0), and its median changed-line ratio is also lower (0.577 vs.\ 0.703). Taken together, these results support a clearer mechanistic claim: SCOPE does not only solve more coder-only failures than Reflexion; it also repairs them more consistently near the same baseline bug signal, while requiring smaller code changes.

SCOPE helps apply compact repairs near the actual fault location when revision is needed. These repairs concentrate on semantic-contract regions rather than on broad structural rewrites.

\section{Discussion}

Our results show that the main value of SCOPE is introducing a prover-initialized critic that makes semantic obligations explicit for revision. This suggests that feedback-based code generation can benefit from structured intermediate reasoning, especially when the main errors are semantic.

\subsection{Implications and Future Work}

\textbf{Reduction of Search Space.}
A key implication of SCOPE is that structured obligations reduce the repair search space. The implementation does not execute Lean tactics, but it borrows the prover-style habit of decomposing a large goal into smaller obligations. Once a programming task is rewritten as concrete obligations, many semantically irrelevant revisions can be ruled out early. The coder is therefore not searching broadly for any program that might pass hidden tests, but revising within a smaller space of edits consistent with the identified obligations. This helps explain why SCOPE tends to produce compact, localized repairs rather than broad rewriting.
As future work, this perspective suggests promising directions. One is to integrate subgoal-guided search more directly into decoding, allowing the coder to plan revisions around partially satisfied obligations instead of relying only on post hoc repair. 

\textbf{Implications for Agentic Coding Systems.}
Our results also suggest a broader implication for agentic software engineering systems. Future coding agents may benefit from maintaining an explicit obligation state between task understanding and code editing. Such a state can help determine which constraint is violated, where the likely fault lies, and whether the next revision should be local or global.
Our findings also provide evidence for role specialization in multi-agent code generation. In SCOPE, a smaller, specialized critic improves a larger coder not by generating better code directly, but by providing a stronger semantic representation of what the draft is missing. This suggests that future agentic systems may be most effective when large coders are paired with smaller expert critics for semantics, localization, and repair planning. 

\subsection{Threats to Validity and Limitations}

Our study has several limitations. First, the evaluation scope is still narrow. SCOPE is trained and tested only on Python code-generation benchmarks, and the tasks are closer to code snippet level programming problems than to full repository scale software engineering workflows. To reduce the risk of overfitting to a single benchmark, we evaluate not only on LiveCodeBench V6 but also on BigCodeBench (Hard), which differs in task composition and execution setup. Even so, broader validation across languages, real repositories, and longer-horizon agentic coding tasks remains future work.

Second, SCOPE is not a full formal verification system. We do not prove that the final Python program satisfies a complete formal specification, and the semantic obligations generated by the critic may themselves be incomplete, approximate, or occasionally misaligned with the true intent of the task. Our goal is therefore more practical than end-to-end certification: we use a prover-initialized model as a structured semantic critic that makes intent more explicit before revision. To reduce the risk that the critic produces merely plausible but unhelpful obligations, we train it with both dense rewards for structured semantic quality and sparse rewards based on downstream execution improvement.

\section{Conclusion}

We present SCOPE, a prover-initialized subgoal critic for Python code generation. By translating natural-language tasks into explicit repair obligations and comparing them against draft programs, SCOPE provides structured feedback that guides targeted repair. Through a three-stage design: execution-grounded data construction, supervised fine-tuning, and process-aligned reinforcement learning with dense semantic reward and sparse execution reward, we show that prover-style subgoal decomposition can be adapted from formal reasoning to practical code generation without claiming full formal verification.

Through evaluation on LiveCodeBench and BigCodeBench, SCOPE achieves the strongest overall performance among the compared baselines, improving over coder-only generation and feedback-based methods such as Reflexion and Self-Refine under the same coder setting. Our analysis further shows that SCOPE helps most on tasks with stronger semantic structure, reduces crash-like failures, and produces more localized repairs rather than broad rewriting. These findings suggest that the main value of the prover-initialized critic in this setting is not end-to-end verification, but making natural-language intent explicit enough to guide reliable revision. 

\bibliographystyle{IEEEtran}
\bibliography{software_fix}
\end{document}